\documentclass[sigconf,screen]{acmart}

\copyrightyear{2021}
\acmYear{2021}
\setcopyright{acmcopyright}
\acmConference[SAC-SVT'21]{The 36th ACM/SIGAPP Symposium on Applied Computing - Software Verification and Testing Track}{March, 2021}{Virtual, Republic of Korea}

\usepackage{graphicx}
\usepackage{comment}
\usepackage{mathtools}
\usepackage{blindtext}
\usepackage{xcolor}
\usepackage{float}
\usepackage{url}
\usepackage{xspace}
\usepackage{multirow}
\usepackage{booktabs}
\usepackage[ruled]{algorithm2e}
\usepackage{pbox}
\usepackage{xspace}
\usepackage{fancybox}
\usepackage{enumitem}
\usepackage{array}
\usepackage{microtype}
\usepackage{balance}

\title{Configuring Test Generators using Bug Reports: A Case Study of GCC Compiler and Csmith}

\author{Md Rafiqul Islam Rabin}
\affiliation{%
   \institution{University of Houston}
   \city{Houston}
   \state{TX}
   \country{USA}}
\email{mrabin@uh.edu}

\author{Mohammad Amin Alipour}
\affiliation{%
   \institution{University of Houston}
   \city{Houston}
   \state{TX}
   \country{USA}}
\email{maalipour@uh.edu}



\begin{document}

\newcommand{\CAST}{ClangAST\xspace}
\newcommand{\CS}{Csmith\xspace}
\newcommand{\KM}{\texttt{K-Means}}
\newcommand{\XM}{X-Means\xspace}

\newcommand{\KC}{\textsc{K-Config}\xspace}
\newcommand{\ksize}{\KC-Weighted\xspace}
\newcommand{\kconfig}{\KC-Round-Robin\xspace}
\newcommand{\swarm}{\textsc{Swarm}}
\newcommand{\default}{\textsc{Default}\xspace}
\newcommand{\regression}{\textsc{Seed}\xspace}
\newcommand{\PSO}{HiCOND\xspace}

\newcommand{\N}{$5,960$\xspace} 
\newcommand{\KMC}{$96$\xspace} 
\newcommand{\KCB}{$112$\xspace} 
\newcommand{\SMC}{$70$\xspace} 
\newcommand{\SCB}{$125$\xspace} 

\newcommand{\test}{test program\xspace}
\newcommand{\Test}{Test program\xspace}
\newcommand{\tests}{test programs\xspace}
\newcommand{\Tests}{Test programs\xspace}
\newcommand{\TS}{$TS$\xspace}
\newcommand{\TG}{\textsc{TestGen}\xspace}

\newcommand{\Fix}[1]{\textbf{\textcolor{red}{Fix/TODO}: #1}}
\newcommand{\Comment}[1]{}
\newcommand{\Space}[1]{}
\newcommand{\Part}[1]{\noindent\textbf{#1}}

\newcounter{observation}
\newcommand{\observation}[1]{\refstepcounter{observation}
        \begin{center}
        \Ovalbox{
        \begin{minipage}{0.93\columnwidth}
                \textbf{Observation \arabic{observation}:} #1
        \end{minipage}
        }
        \end{center}
        \vspace{-5pt}
}

\newcommand{\KCONLY}{3\xspace}
\newcommand{\SWARMONLY}{3\xspace}
\newcommand{\BOTH}{4\xspace}

\begin{abstract}

The correctness of compilers is instrumental in the safety and reliability of other software systems, as bugs in compilers can produce executables that do not reflect the intent of programmers. Such errors are difficult to identify and debug. 
Random test program generators are commonly used in testing compilers, and they have been effective in uncovering bugs. 
However, the problem of guiding these test generators to produce test programs that are more likely to find bugs remains challenging. 

In this paper, we use the code snippets in the bug reports to guide the test generation.
The main idea of this work is to extract insights from the bug reports about the language features that are more prone to inadequate implementation and using the insights to guide the test generators.
We use the GCC C compiler to evaluate the effectiveness of this approach. In particular, we first cluster the test programs in the GCC bugs reports based on their features. We then use the centroids of the clusters to compute configurations for Csmith, a popular test generator for C compilers. 
We evaluated this approach on eight versions of GCC and found that our approach provides higher coverage and triggers more miscompilation failures than the state-of-the-art test generation techniques for GCC.

\end{abstract}

\maketitle

\begin{figure*} [ht]
    \centering
    \includegraphics[width=0.98\linewidth]{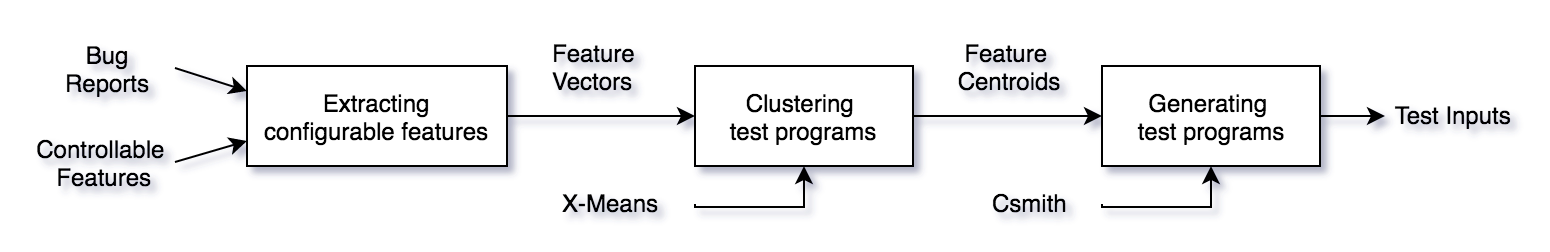}
    \caption{The workflow of test generation in \KC approach.}
    \label{fig:workflow}
\end{figure*}

\begin{figure}
    \centering
    \includegraphics[width=0.92\columnwidth]{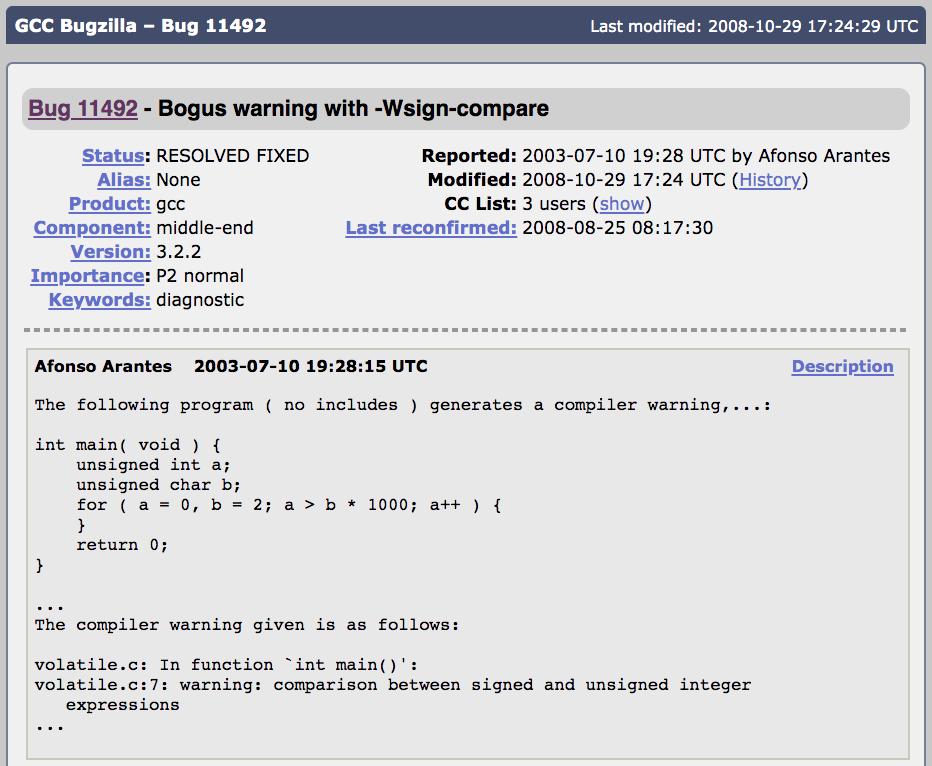}
    \caption{An example of bug report (\#11492) in GCC Bugzilla.}
    \label{fig:bug_report}
\end{figure}

\section{Introduction}

Compilers are key parts of software development infrastructure that translate high-level programs understandable by developers to low-level programs that machines can execute.
Developers \emph{rely} on compilers to build, profile, and debug the programs; therefore any bugs in compilers threaten the integrity of software development process.
For this reason, researchers have been focused on testing compilers to uncover bugs, \cite{Alipour:FocusedTesting:ISSTA:2016,Cummins:TestGenLearn:DeepSmith:ISSTA:2018,Regehr:TestGenGrammar:Csmith:PLDI:2011}. 

Testing compilers is particularly difficult due to their sheer size and complexity. 
Testing such massive, sophisticated systems is a non-trivial task, and researchers and developers still can find bugs in them
\cite{CompilerSurvey:2020}.
Random testing, also know as fuzzing, is a common, lightweight approach for generating \tests for compilers \cite{miller1990empirical}.
Many programming languages have specialized random test generators that use the language grammar and language-specific heuristics to produce test programs in those languages; \CS~\cite{Regehr:TestGenGrammar:Csmith:PLDI:2011} for C, jsFunFuzz~\cite{Link:jsFunFuzz} for JavaScript, and Go-Fuzz~\cite{Link:goFuzz} for Go are good examples of such test generators that have been successful in helping developers to find hundreds of bugs in various compilers and interpreters. 

Mature test generators, like \CS, are configurable and allow developers to guide the test generation through configurations. 
Prior studies have proposed different techniques to exploit configurations to generate more effective test programs.
For example, swarm testing~\cite{Groce:TestGenRnd:Swarm:ISSTA:2012} configures \CS such that test programs contain a random subset of the language features, focused random testing ~\cite{Alipour:FocusedTesting:ISSTA:2016} uses statistical analysis of the programs generated by \CS and their coverage to suggest \CS configurations that target specific blocks in GCC, or
\PSO~\cite{Junjie:PSO:ASE:2019} analyzes on the historical test programs to create \CS configurations that are more likely to find bugs.

In this paper, we propose \KC, an approach that uses the bugs reported by users to propose configurations for \CS.
More specifically, \KC clusters the programs in the bug reports by \XM{} algorithm \cite{Pelleg:Cluster:xmeans:ICML:2000} and uses the centroids of clusters as a basis for proposing configurations for \CS.
We implemented \KC for GCC C compiler and \CS test generator. 
We collected \N{} failing \test{}s from the GCC Bugzilla~\footnote{\label{bugzilla}\url{https://gcc.gnu.org/bugzilla/}} and Testsuite~\footnote{\label{testsuite}\url{https://github.com/gcc-mirror/gcc/blob/master/gcc/testsuite/}}.
We performed an extensive experiment to evaluate the effectiveness of \KC on eight versions of GCC. We compared it with swarm testing~\cite{Groce:TestGenRnd:Swarm:ISSTA:2012}, \PSO~\cite{Junjie:PSO:ASE:2019} and the default configuration of \CS~\cite{Regehr:TestGenGrammar:Csmith:PLDI:2011}.

To the best of our knowledge, the other comparable techniques do not use the information available in the bug reports in the generation of new \tests. \KC is the first attempt in that direction \cite{rabin2019kconfig}. The reasoning it uses is that the code snippets in the bug reports that triggered bugs earlier are more likely to be of interest to developers. We believe bug reports contain insights that can be extracted to improve the test generation for compilers.

The results of our experiments suggest that \KC could find up to \KMC{} miscompilations, while swarm finds up to \SMC{} miscompilations, and the default configurations couldn't find any miscompilations. Moreover, the coverage of \KC is higher than other techniques in all versions of GCC compilers in our experiment.
Our results also show that the coverage of regression \tests is still higher than any automatically generated test suite. It suggests that despite the advances in generating test programs for compilers, there is still a wide gap between the effectiveness of the generated test programs, and the small, regression test suites.

The intuition behind \KC is that the language features that have participated in previous bugs are likely to participate in new bugs as well. 
Similar intuition has been used in LangFuzz~\cite{Zeller:TestGenMutation:LangFuzz:Security:2012} and \PSO~\cite{Junjie:PSO:ASE:2019}.
LangFuzz transplants fragments of failing \test{}s to generate new \tests for JavaScript.
Although similar in the intuition, instead of manipulating existing test programs, \KC uses configurations to guide the test generator to generate new \tests.
\PSO uses the configuration of \CS \emph{generated} test programs to optimize and propose diverse test configurations.
While \KC and \PSO both are similar in the sense that they use an initial test suite to infer effective configurations for \CS, they differ in two main ways. 
First, \PSO requires the configuration of the \CS in the generation of each test program. In other words, all initial test programs must be generated by \CS. Therefore, the test programs that are not created by \CS, like the ones available in the bug reports, cannot be used by \PSO.
Second, the particle swarm optimization technique used in \PSO requires a large initial test suite of failing and passing \tests generated by \CS. Unfortunately, as GCC matures, it becomes harder for \CS to trigger failures in GCC hence, which negatively impacts \PSO applicability on newer, more stable versions of GCC. 

\Part{Contributions}.
This paper makes the following contributions.
\begin{itemize}
    \item We propose an approach for computing configuration of the \CS test generator by processing the code snippets in the bug reports.
    \item We perform a large-scale case study on the effectiveness of our approach on eight versions of GCC---a large, complex C compiler.
\end{itemize}

\Part{Paper Organization}.
The paper is organized as follows.
Section~\ref{sec:approach} describes the proposed \KC approach. 
Section~\ref{sec:experimental} and Section~\ref{sec:evaluation} describe the experimental setting and evaluation setup.
Section~\ref{sec:results} presents the results of the experiments. 
Section~\ref{sec:discussion} discusses the results. 
Section~\ref{sec:related} presents the related works.
Section~\ref{sec:threat} describes the main threats to validity. 
Finally, section~\ref{sec:conclusion} concludes the paper.

\begin{figure*} 
    \centering
    \includegraphics[width=0.98\linewidth]{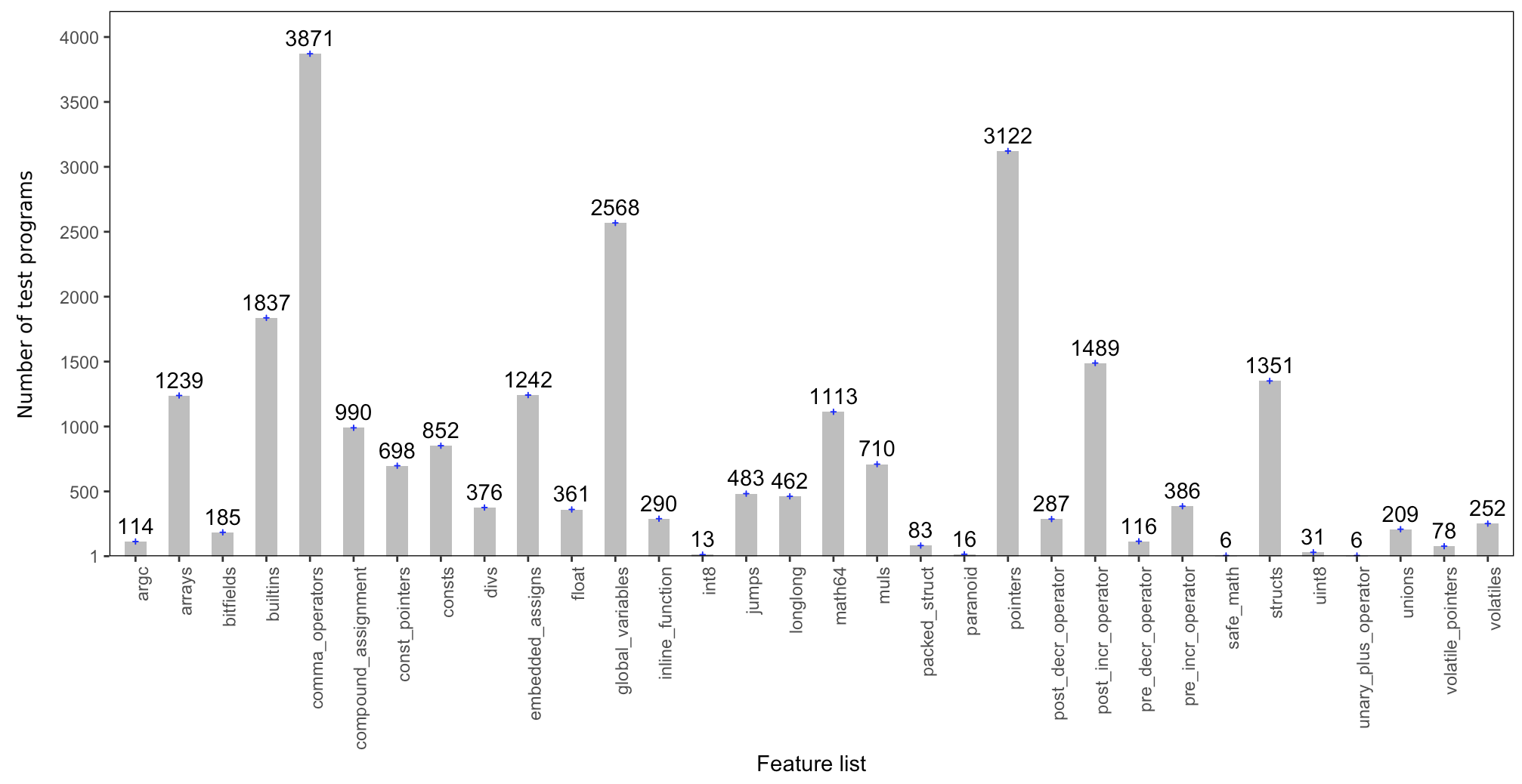}
    \caption{Number of \test{}s in \regression test suite for each feature.}
    \label{fig:feature_count_testinputs}
\end{figure*}

\section{Approach}
\label{sec:approach}

Figure~\ref{fig:workflow} depicts the overall workflow of our approach that constitutes the following three main steps: 
(1) extracting configurable test features from the initial test suite, 
(2) using the test features to cluster \test{}s into similar groups, and
(3) generating configurations based on the centroid of clusters.
We describe each step in the rest of this section.

\subsection{Extracting configurable test features}

\Part{Test features.}
We use the term \emph{test features}, or \emph{features} for short, similar to Groce et al.~\cite{Groce:ISSRE:2013}: ``Test features are basic \emph{compositional units} used to describe what a test does (and does not) involve.'' Test features are essentially building blocks of test cases. For example, in testing libraries, the API calls are the functions; in grammar-based testing, the production rules in the grammar can be features; and, in testing compilers or interpreters, programming structures can be test features. In \CS, test features are the same as C language features, therefore, in the rest of the paper, we use terms test feature, language feature, and feature interchangeably.

\Part{Configurable test generator.}
A test generator essentially composes new \tests with or based on the test features. 
Some test generators allow users to use configuration files or command-line options to customize test generation by modifying the test generation parameters. 
We call such test generators \emph{configurable}.
\CS{}~\cite{Regehr:TestGenGrammar:Csmith:PLDI:2011} is a configurable test generator for C compilers. 
\CS{} then uses these parameters to decide which features to include in the test programs.

\Part{Controllable test features in \CS.}
\CS{} allows choosing the C programming constructs to be included in the \tests through command-line options.
The order and number of the constructs are, however, chosen randomly and developers cannot control them---mainly because \CS{} is a random generator that uses grammar to generate \test{}s. 
As of \CS{} version $2.3.0$ \footnote{\label{csmith}\url{https://embed.cs.utah.edu/csmith/}}, the \CS{} provides $32$ configurable features in the form of \texttt{--ft} and \texttt{--no-ft} command-line options, where $ft$ denotes a feature, and \texttt{--ft} indicates inclusion and \texttt{--no-ft} indicates exclusion of $ft$ in the generation of the \tests. 
Figure ~\ref{fig:feature_count_testinputs} shows the list of available controllable features in \CS. 
These features describe a wide range of language features from programming constructs in C language, e.g., \texttt{compound\_assignment} for compound assignments, and \texttt{float} for floating-point data types.
Different features may impact the behavior of the compiler under test in different ways. 
For example, Groce et al.~\cite{Groce:ISSRE:2013} observed that pointer manipulation in a \test can suppress certain class of bugs in the C compilers, as most compilers would conservatively avoid certain optimizations for pointers in the input program, consequently, masking potential bugs in those optimizations.

\Part{Extracting features.}
In the context of this paper, we only consider \tests that have been reported in the GCC bug reports.
Figure ~\ref{fig:bug_report} shows an example of bug report (\#11492) in GCC Bugzilla \footnote{\url{https://gcc.gnu.org/bugzilla/show_bug.cgi?id=11492}}.
We use the number of occurrences of features in each \test to create a feature vector for the \test.
For this, we use \CAST{} \footnote{\url{https://clang.llvm.org/docs/IntroductionToTheClangAST.html}} to extract the abstract syntax tree (AST) of the programs in the initial failing test suite.
Feature vectors are normalized to the range of $[0,1]$ using min-max normalization~\cite{Pedregosa:ScikitLearn:2011}, i.e.,
$ z_{ij} = \frac { x_{ij} \, - \, \min(x_{.j}) } { \max(x_{.j}) \, - \, \min(x_{.j}) } $, where $x_{ij}$ is the value of the $j^{th}$ feature of the $i^{th}$ \test $x_{i}=(x_{i1},...,x_{in})$, $x_{.j}$ is the list of $j^{th}$ feature from all \tests, and $z_{ij}$ is the corresponding min-max normalized value of $x_{ij}$.

\subsection{Clustering \tests}
To identify the clusters, we use \XM{} clustering algorithm \cite{Pelleg:Cluster:xmeans:ICML:2000} that estimates the optimal number of clusters in the underlying distribution of data.
The \XM{} clustering is an unsupervised machine learning algorithm that performs clustering of unlabeled data without the need for presetting the number of clusters.
Each cluster is represented by a \emph{centroid} that has a minimum distance to the data points of the cluster.
Since the feature vectors contain the value in the range of $[0,1]$, the values in the centroids would be a number between $0$ and $1$ as well.

\subsection{Generating \tests}
Our implementation of \KC uses the centroid in \XM{} clustering \cite{Pelleg:Cluster:xmeans:ICML:2000} to suggest configurations for the test generator.
To generate configurations, we use the corresponding value of a feature in the cluster as the probability of including the feature in a \test{}.
The feature is more dominant in the \test{}s if the corresponding value of a feature is closer to $1$, and conversely, if the corresponding value is closer to $0$, it denotes that the feature is less prevalent in the \test{}s.
Therefore, in this approach, for each cluster, we create a configuration wherein the probability of inclusion of a feature in the \test is equal to its corresponding value in the centroid of that cluster.
For instance, suppose the centroid of a cluster is $(0.1, 0.7)$ for features \texttt{f1} and \texttt{f2}, respectively, 
the configuration generation algorithm in \KC, whenever called, it includes \texttt{--f1} with probability $0.1$ and \texttt{--no-f1} with probability $0.9$, similarly it would \texttt{--f2} with $0.7$ probability and \texttt{--no-f2} with $0.3$ probability.
It is in contrast with the swarm testing~\cite{Groce:TestGenRnd:Swarm:ISSTA:2012} that uses the simplest form of fair coin-toss probability (i.e., $0.5$) for the inclusion of a feature.

Algorithm~\ref{alg:generator} describes the process of generating new \test{}s using \KC in more detail.
Given a testing budget $totalBudget$ and a set of centroids $CS$, the algorithm calls $ConfigGen$ in round-robin fashion until the test budget expires.  
Procedure $ConfigGen$ takes a centroid $C\in CS$ and generates a new configuration.
In generating a new configuration, $ConfigGen$ chooses to include feature $f_i$ with a random probability $c_i$ where $f_i$ is represented by the element $c_i$ in the centroid $C$. 
Finally, $TS$ will have all the generated failure-inducing test programs.

\begin{algorithm} [ht]
\caption{\KC}
\label{alg:generator}
 $totalBudget$  $\leftarrow$ Testing budget (i.e. 10,000 test count)\;
 $CS$ $\leftarrow$ Set of centroids\;
 $TS \leftarrow \{\}$\; 
 \DontPrintSemicolon\;
 
 \While{$spentBudget \leq totalBudget$}{
 \ForAll{centroid $C$ $\in$ $CS$}{
     $config \leftarrow ConfigGen(C)$\;
     $test \leftarrow Csmith(config)$\;
     \If{doesFail(test,GCC)}{
      $TS \leftarrow TS \cup test$\;
     }
    }
}
     
\DontPrintSemicolon\;
\SetKwProg{Fn}{Function}{:}{\KwRet}
\Fn{ConfigGen($C$)}{
    $features \leftarrow \emptyset$\;
    \ForAll{value $c$ $\in$ $C$}{
      $randProb \leftarrow [0:1]$\;
    \eIf{randProb $\leq$ c}{
        $features.put(1)$\;
    }{
        $features.put(0)$\;
    }
 }
}
\end{algorithm}

\section{Experimental Setting}
\label{sec:experimental}
This section discusses the experimental setting we used for the \KC approach.

\Part{Reference test suite}.
A reference test suite is a collection of \tests that exhibit interesting behaviors, e.g., high code coverage or triggering failures.
We use the GCC regression test suite that has \N{} parsable C code snippets.
These code snippets have been collected from the confirmed bug reports in GCC Bugzilla $^{\ref{bugzilla}}$ and Testsuite $^{\ref{testsuite}}$.
The code snippets in the bug reports are small and usually do not include \texttt{main} functions.
We call this test suite \regression henceforth.
We, hereby, use terms regression test suite and \regression test suite interchangeably.

Figure ~\ref{fig:feature_count_testinputs} shows the number of \test{}s in \regression test suite for each feature.
It shows that the distribution of test features in failing \test{}s is not uniform. 
Features such as comma operators, global variables, and pointers occurred more frequently than features such as unary plus operator (\texttt{+}), safe math, or int8. 
We observe that all \CS{} configurable test features appear in one or more failing \tests in the regression test suite. Each feature has been present in $6$ to $3871$ \tests.
Note that it is difficult to identify the role of individual features in the failures in the large, complex systems such as GCC, without substantial simplification of the test~\cite{DD} and close inspection of the program execution.

\Part{Test generation tool}.
We use \CS{} \cite{Regehr:TestGenGrammar:Csmith:PLDI:2011} that is an open-source automatic test generation tool for C compilers. 
Given a set of C language features as options, \CS{} can generate a random C program that contains those features. 
The programs generated by \CS{} are \emph{closed}; that is, all variables are initialized in the source code by \CS and they do not require any external inputs for the execution. 
We use \CS{} $2.3.0$ $^{\ref{csmith}}$ in our experiments.

\Part{Test generation techniques}.
We compare the effectiveness of configurations proposed by \KC with the effectiveness of configurations in three comparable techniques: (1) \CS default configuration, (2) swarm configurations, and (3) \PSO configurations. 
In the \emph{default configuration}, we use the default configuration of \CS wherein all C language features are enabled by default.
\emph{Swarm configurations} are configurations that the \CS features are enabled by a fair coin-toss as in ~\cite{Groce:TestGenRnd:Swarm:ISSTA:2012}.
\emph{\PSO configurations} are created by particle swarm optimization as in ~\cite{Junjie:PSO:ASE:2019} that attempts to optimize the effectiveness of \CS by systematically exploring the bug-finding capability of configurations in \CS.

\Part{Configuration generation strategies}.
We evaluate \KC with two configuration generation strategies: round-robin, and weighted.
In the round-robin strategy, centroids of the clusters are used in a round-robin fashion to generate \tests.
Round-robin strategy ignores the size of clusters.
The weighted \KC strategy is similar to the round-robin, except the number of \tests generated by each centroid is proportional to the size of the cluster that the centroid represents. 
In weighted \KC, larger clusters will have more \tests generated.

\Part{GCC versions under test}.
We use eight versions of GCC to evaluate the effectiveness of the \KC approach: GCC $4.3.0$, $4.8.2$, $5.4.0$, $6.1.0$, $7.1.0$, $8.1.0$, $9.1.0$, and GCC trunk (as of $09/03/2019$).
The official releases are mature and have been widely in use for building various programs and operating systems,
while the trunk version contains the newest experimental features and, as a result, is not as stable as the official releases. 

\section{Evaluation Setup}
\label{sec:evaluation}

\begin{table}
    \begin{center}
        \def\arraystretch{1.2}
        \resizebox{\columnwidth}{!}{%
        \begin{tabular}{|c|c|c|}
            \hline
            \textbf{No Optimization (-O0) } & \textbf{High Optimization (-O3)} & \textbf{Failure?} \\
            \hline \hline 
                Compiler crashes & Compiler crashes & False \\ \hline
                Compiler crashes & Compiler doesn't crash & True \\ \hline
                Compiler doesn't crash & Compiler crashes & True \\ \hline
                \multicolumn{2}{|c|} {Outputs are identical for different optimization} & False \\ \hline
                \multicolumn{2}{|c|} {Outputs are different for different optimization} & True \\ 
            \hline
        \end{tabular}%
        }
        \vspace{2mm}
        \caption{Possible failures in the experiment.}
        \label{table:problematic_combinations}
        \vspace{-2mm}
    \end{center}
\end{table}

\begin{figure}
    \centering
    \includegraphics[width=0.85\columnwidth]{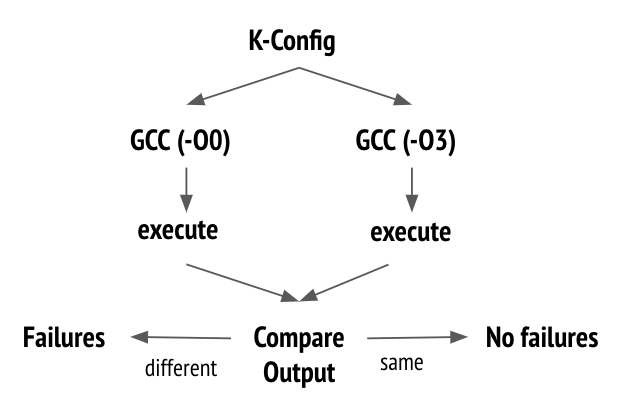}
    \caption{Differential testing in optimization level.}
    \label{fig:diff_test}
\end{figure}

This section discusses the evaluation setup we used for the \KC approach.

\Part{Test oracles}.
The common practice for testing compilers is differential testing \cite{mckeeman1998differential, Regehr:TestGenGrammar:Csmith:PLDI:2011}.
That is, a \test{} is compiled and executed by two or more versions of compilers, or two or more optimization levels, and the results are compared.
The differential test oracle specifies the result of the output of the compiled programs by all compilers and optimization levels must be the same.
Undefined behaviors \cite{Regehr:TestGenGrammar:Csmith:PLDI:2011} in the C language can complicate this process, however, \CS{} does the best effort to avoid the undefined behaviors in C.

We use differential testing to evaluate the behavior of the compilers on the \tests.
Figure ~\ref{fig:diff_test} shows the differential testing in optimization level.
More specifically, we compile a program with different levels of optimization and compare the output of the executed programs. 
Due to the large scale of the experiment, and to optimize the experimentation time, we only use the lowest \texttt{-O0} and highest \texttt{-O3} optimization levels. We do not include the intermediate optimization levels \texttt{-O1} and \texttt{-O2}, as most optimization passes that a compilation of a program can invoke can be captured in \texttt{-O3}.
Note that although almost all optimization passes are enabled in \texttt{-O3}, there may be cases that the interactions of viable, competing optimizations may put compiler in a position to choose ones over another one, hence not exercising some of the passes that would have been otherwise exercised in \texttt{-O1} or \texttt{-O2} levels. Our experiment does not capture such cases, and we speculate that such cases would be rare and hence negligible.

\Part{Failure types}.
We expect three types of compiler failures in the compilation of a program: (1) Crash when compiler terminates the compilation abruptly with a crash report on screen, (2) Timeout when compiler gets into non-termination on the compilation of a program, and (3) Miscompilation when the executable generated by the compiler produced different results on different optimizations.
We check the behavior of the compiler (i.e., exit code) to identify the crashes.
We use $10$ seconds timeout in compiling \tests.
We identify miscompilation by comparing the output of executables generated by the compiler on different optimization levels.

\Part{Experiment parameters}.
Overall we evaluate five configuration generation for \CS{}: \default, \swarm, \PSO, \kconfig and \ksize.
\default{} denotes the default configuration that includes all the features by default.
\swarm{} represents the swarm testing configuration where the probability of inclusion of a feature in a \test is $0.5$, same as the fair coin-toss approach \cite{Groce:TestGenRnd:Swarm:ISSTA:2012}.
\PSO denotes the test configurations computed from the history-guided approach proposed by Chen et al.~\cite{Junjie:PSO:ASE:2019}.
In \KC, this probability is equal to the corresponding value in the centroid of the cluster.
For example, if the corresponding value for \texttt{volatile} in the centroid of a cluster is $0.2$, there is $20\%$ chance that \KC to use \texttt{--volatile} and $80\%$ chance that it includes \texttt{--no-volatile} in the configuration.
\kconfig{} adopts the round-robin strategy with uniform weights for all clusters, while \ksize{} uses a weighted strategy based on the size of clusters.
Note that we do not evaluate the regression test suite, as only a few of its \tests have \texttt{main} functions that can generate an executable.

\Part{Clustering parameters}.
We use the implementation of \XM{} algorithm in \texttt{pyclustring} data mining library ~\cite{Novikov:PyClustering:2019} with default hyperparameters.
We use the default Bayesian information criterion (BIC)~\cite{BIC} for the splitting type to approximate the number of clusters.
The stop condition for each iteration is $0.025$;
the algorithm stops processing whenever the maximum value of change in centers of clusters is less than this value.
Moreover, the default distance metric used is the sum of squared errors (SSE) \cite{kwedlo2011clustering} which is the distance between data points and its centroid.

\Part{Test budget}.
We create $10,000$ \tests for each approach and evaluate the effectiveness of the test (i.e., bug finding, coverage, and the number of distinct bugs in each approach).
To account for the random effects in the approaches, we run each experiment three times, average them, and round the values to the closest integer.
We use $10$ seconds as the timeout for GCC to compile a \test{}.
A small experiment with 6 hours test budget and 30 seconds timeout yielded similar observations.
Overall, we evaluated the techniques for over $1,500$ hours (approximately $62$ days).

\section{Results}
\label{sec:results}

\newcommand{\VR}{VR\xspace}
\newcommand{\ST}{ST\xspace}
\newcommand{\TI}{TI\xspace}
\newcommand{\Cz}{C0\xspace}
\newcommand{\Cth}{C3\xspace}
\newcommand{\Czth}{C03\xspace}
\newcommand{\TC}{TC\xspace}
\newcommand{\Tz}{T0\xspace}
\newcommand{\Tth}{T3\xspace}
\newcommand{\Tzth}{T03\xspace}
\newcommand{\TT}{TT\xspace}
\newcommand{\MC}{MC\xspace}

In this section, we present the results of our experiments.
Table \ref{table:result_10k} depicts the number of failures in each experiment. 
In this table, the column \Cz and \Cth denote the number of \tests that crashed \emph{only} in -O0 and \emph{only} in -O3, respectively, and \Czth denotes the number of \tests that crashed in \emph{both}. 
Similarly, the column \Tz, \Tth, and \Tzth represent the number of \tests that encountered timeouts \emph{only} in -O0, \emph{only} in -O3, and in \emph{both}, respectively.
Finally, the column \MC indicates the total number of \tests that produced miscompilations.

\newcommand{\DESC}{
    \VR: GCC Version, 
    \ST: Setting, 
    \TI: \#\tests, 
    \Cz: \#crashes under -O0, 
    \Cth: \#crashes under -O3, 
    \Czth: \#crashes under both -O0 and -O3, 
    \TC: \# of all crashes,
    \Tz: \#timeouts in -O0, 
    \Tth: \#timeouts in -O3,
    \Tzth: \#timeouts in both -O0 and -O3,
    \TT: \# of all timeouts,
    \MC: \#miscompilations
}

\begin{table}
    \centering
    \def\arraystretch{1.1}
    \resizebox{0.99\columnwidth}{!}{%
    \begin{tabular}{|c|c|c|c|c|c|c|c|c|}
        \hline
        \textbf{GCC} & \textbf{Setting} & \textbf{\Cz} & \textbf{\Cth} & \textbf{\Czth} & \textbf{\Tz} & \textbf{\Tth} & \textbf{\Tzth} & \textbf{\MC} \\   
        
        \hline \hline
        
        \multirow{5}{*}{4.3.0}
        &\default & 0 & 1708 & 0 & 242 & 0 & 1017 & 16 \\
        \cline{2-9}
        &\swarm & \textbf{256} & 349 & \textbf{61} & 1159 & 63 & 1374 & 55 \\
        \cline{2-9}
        & \PSO & 0 & \textbf{2584} & 0 & 386 & 1 & 990 & 16 \\
        \cline{2-9}
        &\kconfig & 187 & 519 & 49 & 1256 & 45 & 1101 & \textbf{68} \\
        \cline{2-9}
        &\ksize & 133 & 329 & 20 & 2003 & 33 & 947 & 29 \\
        \hline \hline
        
        \multirow{5}{*}{4.8.2}
        &\default & 0 & 0 & 0 & 10 & 0 & 1250 & 0 \\
        \cline{2-9}
        &\swarm & 56 & 65 & 4 & 1218 & 18 & 1410 & 52 \\
        \cline{2-9}
        & \PSO & 0 & 0 & 0 & 9 & 0 & 1366 & 0 \\
        \cline{2-9}
        &\kconfig & 45 & 61 & 5 & 1269 & 16 & 1198 & \textbf{73} \\
        \cline{2-9}
        &\ksize & 26 & 35 & 2 & 2266 & 4 & 993 & 21 \\
        \hline \hline
        
        \multirow{4}{*}{5.4.0}
        &\default & 0 & 0 & 0 & 11 & 0 & 1249 & 0 \\
        \cline{2-9}
        &\swarm & 0 & 0 & 0 & 1224 & 5 & 1429 & 61 \\
        \cline{2-9}
        &\kconfig & 0 & 0 & 0 & 1276 & 6 & 1217 & \textbf{81} \\
        \cline{2-9}
        &\ksize & 0 & 0 & 0 & 2003 & 0 & 973 & 25 \\
        \hline \hline
        
        \multirow{4}{*}{6.1.0}
        &\default & 0 & 0 & 0 & 13 & 0 & 1247 & 0 \\
        \cline{2-9}
        &\swarm & 0 & 0 & 0 & 1253 & 4 & 1396 & 64 \\
        \cline{2-9}
        &\kconfig & 0 & 0 & 0 & 1298 & 6 & 1188 & \textbf{82} \\
        \cline{2-9}
        &\ksize & 0 & 0 & 0 & 2005 & 0 & 961 & 26 \\
        \hline \hline
        
        \multirow{4}{*}{7.1.0}
        &\default & 0 & 0 & 0 & 12 & 0 & 1247 & 0 \\
        \cline{2-9}
        &\swarm & 0 & 0 & 0 & 1135 & 6 & 1490 & 66 \\
        \cline{2-9}
        &\kconfig & 0 & 0 & 0 & 1193 & 9 & 1270 & \textbf{86} \\
        \cline{2-9}
        &\ksize & 0 & 0 & 0 & 1939 & 4 & 1018 & 28 \\
        \hline \hline
        
        \multirow{4}{*}{8.1.0}
        &\default & 0 & 0 & 0 & 13 & 0 & 1247 & 0 \\
        \cline{2-9}
        &\swarm & 0 & 0 & 0 & 1132 & 6 & 1478 & 66 \\
        \cline{2-9}
        &\kconfig & 0 & 0 & 0 & 1190 & 8 & 1269 & \textbf{88} \\
        \cline{2-9}
        &\ksize & 0 & 0 & 0 & 1939 & 4 & 1017 & 27 \\
        \hline \hline
        
        \multirow{4}{*}{9.1.0}
        &\default & 0 & 0 & 0 & 13 & 0 & 1247 & 0 \\
        \cline{2-9}
        &\swarm & 0 & 0 & 0 & 1143 & 6 & 1468 & 69 \\
        \cline{2-9}
        &\kconfig & 0 & 0 & 0 & 1209 & 8 & 1250 & \textbf{95} \\
        \cline{2-9}
        &\ksize & 0 & 0 & 0 & 1965 & 4 & 991 & 26 \\
        \hline \hline
        
        \multirow{4}{*}{trunk}
        &\default & 0 & 0 & 0 & 45 & 0 & 1215 & 0 \\
        \cline{2-9}
        &\swarm & 0 & 0 & 0 & 1410 & 5 & 1200 & 70 \\
        \cline{2-9}
        &\kconfig & 0 & 0 & 0 & 1359 & 7 & 1099 & \textbf{96} \\
        \cline{2-9}
        &\ksize & 0 & 0 & 0 & 2150 & 4 & 805 & 26 \\
        \hline
    \end{tabular}%
    }
    \vspace{2mm}
    \caption{Result for 10,000 \tests.}
    \label{table:result_10k}
    \vspace{-4mm}
\end{table}

\subsection{Comparison with \default and \swarm}

In Table \ref{table:result_10k}, $10,000$ \test{}s of \kconfig{}, on average, triggered up to $96$ miscompilations, while \swarm{} triggered up to $70$ miscompilations, on average. The default configuration of \CS{} (\default) triggered $16$ miscompilations in GCC-4.3.0 but did not find any miscompilations in other versions of GCC.

The \default found $1708$ crashes in GCC-4.3.0 compared to $755$ by \kconfig{} and $666$ by \swarm{}. However, the \default did not find any crashes in GCC-4.8.2, where the \kconfig{} found $111$ crashes and the \swarm{} found $125$ crashes.

In all cases, \kconfig{} triggered the highest number of miscompilations compared to \default and \swarm.

\subsection{Comparison with \PSO}

\PSO uses particle swarm optimization to search for configurations that can find more bugs. 
We received the configurations of \PSO for GCC-4.3.0 from the authors.
In GCC-4.3.0, \PSO finds the highest number of crashes on \texttt{-O3}, while swarm followed by \kconfig{} discover the highest number of crashes under \texttt{-O1}. \kconfig{} finds the highest number of miscompilations.  
Configurations proposed by \PSO do not find any crashes in GCC-4.8.2, as it is highly optimized for bugs in GCC-4.3.0. 
Moreover, \PSO relies on a set of bugs generated by \default to search for the optimal configurations. 
\default does not trigger any failures on GCC-4.8.2, we, therefore, could not compute new configurations using the approach in \cite{Junjie:PSO:ASE:2019} for GCC-4.8.2 and beyond.

\subsection{Number of distinct bugs}

To identify the number of distinct bugs we used the correcting commits heuristic that has been used in previous studies~\cite{Regehr:TestSelection:Rank:PLDI:2013, Chen:TestSelection:LET:ICSE:2017}.
This heuristic uses the commit that a test program switched from failing to passing as the proxy for the minimum number of distinct bugs.
Note that in our results, there are test programs that still exhibit miscompilation characteristics on the latest versions of the GCC, therefore, this heuristic can not be used for them.
Due to the submission deadline, we only compute the distinct bugs for \swarm{} and \kconfig in GCC-4.8.2.
For the crash and miscompilation of failing \tests, the numbers of distinct bugs detected by the techniques are: \KCONLY only by \kconfig, \SWARMONLY only by \swarm, and \BOTH by both.

\subsection{Effectiveness of cluster weighting strategies}

We evaluated the effectiveness of \KC for two cluster weighting strategies: \kconfig{} that ignores the size of clusters, and \ksize{} which generates test programs proportional to the size of clusters.
The results in Table \ref{table:result_10k} show that \kconfig{} was more effective than \ksize{} in finding bugs. 
\kconfig{} could trigger twice as more miscompilations, and thrice as more crashes than \ksize{}. 
The potential reason can be that larger clusters represent prominent combinations of features in the bug reports, it is likely that developers already noticed them and addressed them. Therefore, \tests generated based on \ksize{} would not lead to new failures.
Due to the poor performance of \ksize, we excluded the \ksize{} from subsequent experiments to save computing time.

\subsection{Effectiveness of individual clusters}

\begin{figure}
    \centering
    \includegraphics[width=\columnwidth]{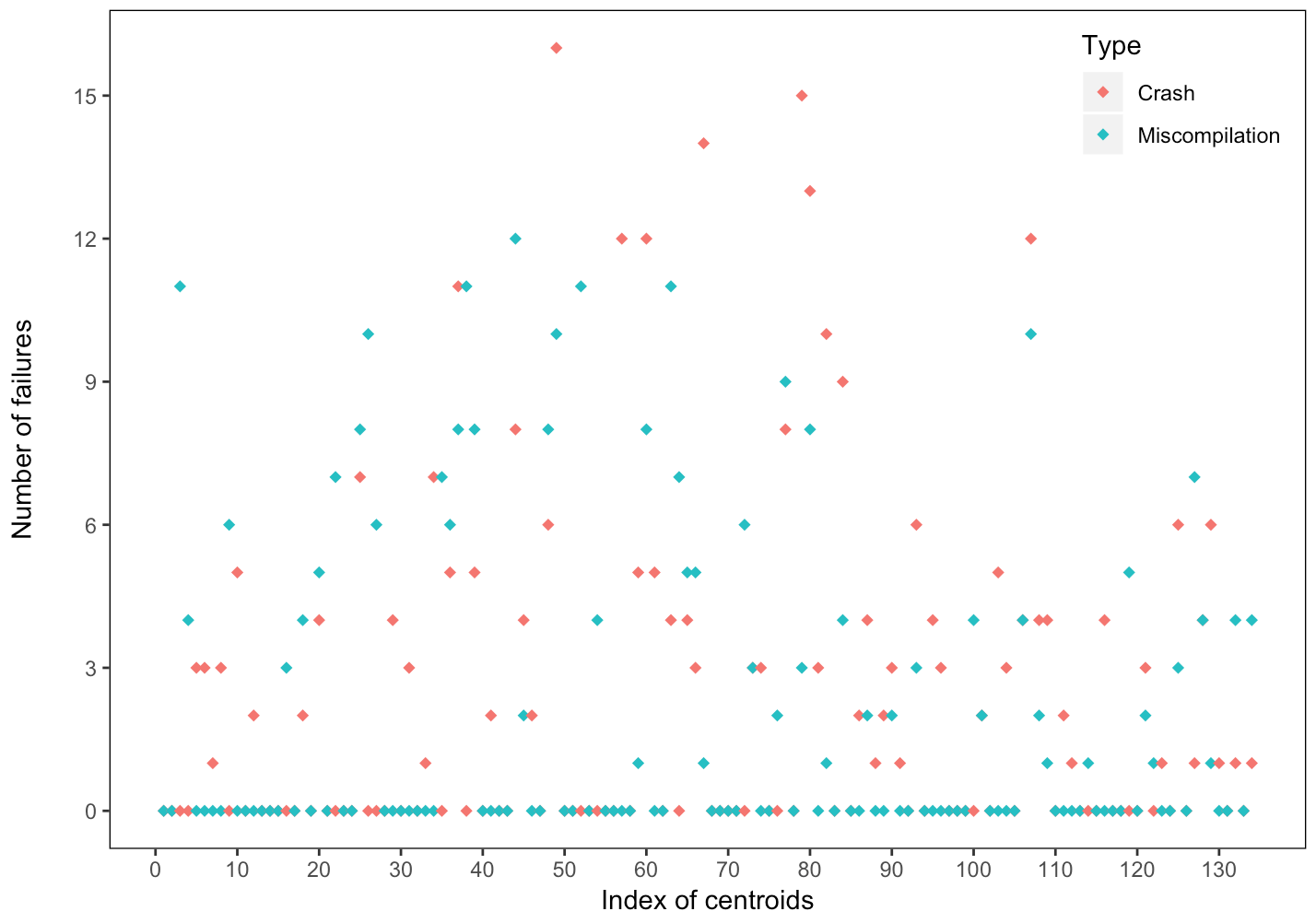}
    \caption{Number of failures in each configuration.}
    \label{fig:nbug}
\end{figure}

To measure the effectiveness of individual clusters in \kconfig, we count the number of failures triggered by the \tests generated based on their centroid configuration.
Figure ~\ref{fig:nbug} shows the number of crashes and miscompilations triggered by each cluster. We excluded the timeouts from the figure due to their sheer numbers.
The $x$-axis in this figure denotes the clusters, and the $y$-axis denotes the number of miscompilation and crash failures triggered by each cluster.

Among $134$ clusters, $50$ did not contribute to finding any failures, $29$ triggered only one crash, $18$ triggered only one miscompilation, while there are configurations with $20$ to $26$ (crash or miscompilation) failures.
Table~\ref{table:top5_nbug} shows the top-$5$ most effective configurations in triggering failures (crash and miscompilation combined).

\begin{table}
    \begin{center}
        \def\arraystretch{1.35}
        \resizebox{\columnwidth}{!}{%
        \begin{tabular}{|c|p{5.8cm}|c|c|c|}
            \hline
            \textbf{Index} & \textbf{Features in Centroid} & \textbf{\# Crash} & \textbf{\# MC} & \textbf{\# Total} \\ \hline
            49 & 
            {
                arrays, comma-operators, compound-assignment, embedded-assigns, 
                pre-decr-operator, post-incr-operator, unary-plus-operator, longlong,
                float, math64, muls, volatiles, volatile-pointers, global-variabless, builtins
            }
            & 16 & 10 & 26 \\ \hline
            107 & 
            {
                bitfields, consts, divs, embedded-assigns, pre-incr-operator, 
                pre-decr-operator, post-decr-operator, unary-plus-operator, jumps, 
                longlong, int8, uint8, float, math64, muls, safe-math, packed-struct, 
                paranoid, structs, unions, volatiles, volatile-pointers, const-pointers, 
                global-variabless, builtins
            }
            & 12 & 10 & 22 \\ \hline
            80 & 
            {
                argc, arrays, bitfields, compound-assignment, consts, divs, embedded-assigns, 
                pre-incr-operator, pre-decr-operator, post-incr-operator, post-decr-operator, 
                unary-plus-operator, jumps, int8, uint8, float, inline-function, muls, safe-math, 
                packed-struct, paranoid, structs, unions, volatiles, volatile-pointers, 
                const-pointers, global-variabless, builtins
            }
            & 13 & 8 & 21\\ \hline
            60 & 
            {
                argc, arrays, bitfields, comma-operators, compound-assignment, divs, 
                embedded-assigns, post-incr-operator, post-decr-operator, 
                unary-plus-operator, uint8, float, math64, inline-function, muls, 
                packed-struct, structs, unions, volatiles, volatile-pointers, 
                global-variabless, builtins
            }
            & 12 & 8 & 20\\ \hline
            44 & 
            {
                comma-operators, compound-assignment, embedded-assigns, unions, 
                volatiles, global-variabless, builtins
            }
            & 8 & 12 & 20\\ \hline
        \end{tabular}%
        }
        $\newline$
        \caption{Top-5 centroids for failure-inducing \test{}s}
        \label{table:top5_nbug}
    \end{center}
\end{table}

\subsection{Coverage across test suites}

Table~\ref{table:coverage_10k} shows the number of covered statements, branches, and functions on GCC versions 4.3.0, 4.8.2, 5.4.0, 6.1.0, and 7.1.0 for the \regression{}, \default{}, \swarm{}, and \kconfig{} test suites.
We computed the coverage with -O3 optimization. 
At the time of writing, we did not extract the coverage information for most recent versions of GCC, i.e., 8.1.0, 9.1.0, and trunk, due to difficulties stemming from recent changes in the file structure in GCC project.
The coverage shows that, in the measured GCC versions, the regression test suite has higher coverage than the generated test suites. 
In the generated test suite, the \kconfig{} has higher coverage than others (\default{}, \swarm{}, and \PSO{}).

\begin{table}
  \begin{center}
    \def\arraystretch{1.35}
    \resizebox{\columnwidth}{!}{%
    \begin{tabular}{|c|c|l|l|l|}
        \hline
        \textbf{GCC} & \textbf{Setting} & \textbf{\# Statement} & \textbf{\# Branch} & \textbf{\# Function} \\   
        \hline \hline
        
        \multirow{5}{*}{4.3.0}
        & \regression & 169838\checkmark\checkmark & 182429\checkmark\checkmark & 10933\checkmark\checkmark \\
        \cline{2-5}
        & \default & 133700 & 138581 & \phantom{0}9181 \\
        \cline{2-5}
        & \swarm & 140404 & 148677 & \phantom{0}9475 \\
        \cline{2-5}
        & \PSO & 134721 & 139570 & \phantom{0}9252 \\
        \cline{2-5}
        & \kconfig & 141165\checkmark & 148910\checkmark & \phantom{0}9499\checkmark \\
        \hline \hline
        
        \multirow{5}{*}{4.8.2}
        & \regression & 214054\checkmark\checkmark & 210324\checkmark\checkmark & 17601\checkmark\checkmark \\
        \cline{2-5}
        & \default & 177269 & 171935 & 15464 \\
        \cline{2-5}
        & \swarm & 183123 & 180203 & 15688 \\
        \cline{2-5}
        & \PSO & 176646 & 171637 & 15460 \\
        \cline{2-5}
        & \kconfig & 183736\checkmark & 180686\checkmark & 15734\checkmark \\
        \hline \hline
        
        \multirow{4}{*}{5.4.0}
        & \regression & 256300\checkmark\checkmark & 251049\checkmark\checkmark & 21949\checkmark\checkmark \\
        \cline{2-5}
        & \default & 207342 & 199413 & 19088 \\
        \cline{2-5}
        & \swarm & 216008 & 210644 & 19431 \\
        \cline{2-5}
        & \kconfig & 216438\checkmark & 211039\checkmark & 19436\checkmark \\
        \hline \hline
        
        \multirow{4}{*}{6.1.0}
        & \regression & 275888\checkmark\checkmark & 269575\checkmark\checkmark & 23865\checkmark\checkmark \\
        \cline{2-5}
        & \default & 222485 & 214760 & 20398 \\
        \cline{2-5}
        & \swarm & 232007 & 226936 & 20848 \\
        \cline{2-5}
        & \kconfig & 232990\checkmark & 227865\checkmark & 20899\checkmark \\
        \hline \hline
        
        \multirow{4}{*}{7.1.0}
        & \regression & 294633\checkmark\checkmark & 291922\checkmark\checkmark & 25131\checkmark\checkmark \\
        \cline{2-5}
        & \default & 237252 & 233134 & 21389 \\
        \cline{2-5}
        & \swarm & 248171 & 247166 & 21953\checkmark \\
        \cline{2-5}
        & \kconfig & 248654\checkmark & 247621\checkmark & 21953\checkmark \\
        \hline
    \end{tabular}%
    }
    $\newline$
    \caption{Number of statements, branches, and functions covered in test suites. (\checkmark\checkmark denotes the test suite with the highest coverage. \checkmark denotes the test suite with the second highest coverage.)}
    \label{table:coverage_10k}
  \end{center}
\end{table}

\subsection{The probability of features}

\begin{figure}
    \centering
    \includegraphics[width=\columnwidth]{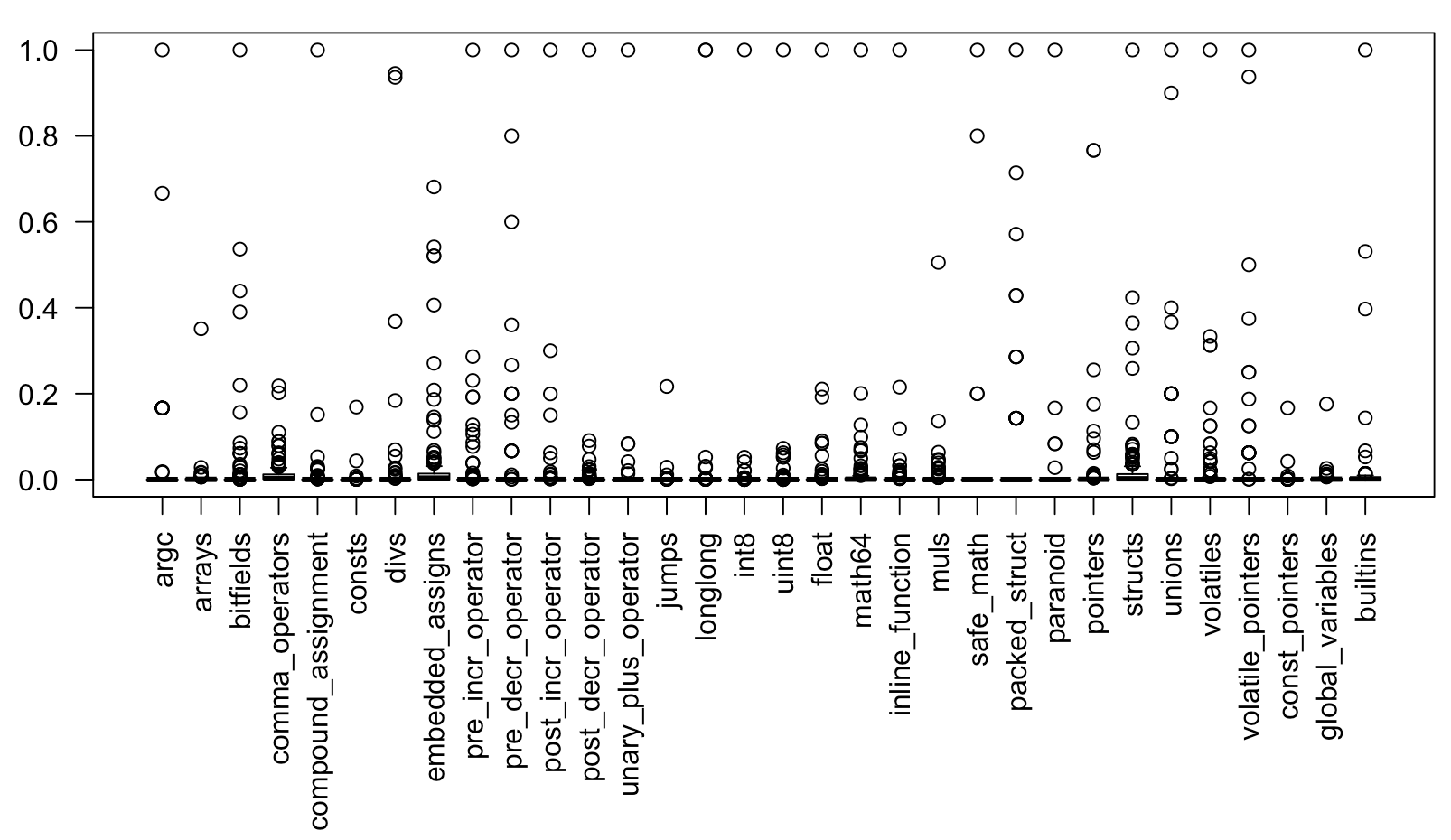}
    \caption{Distribution of probability of inclusion for individual GCC features in \kconfig.}
    \label{fig:centroid:probabilities}
\end{figure}

In the swarm testing~\cite{Groce:TestGenRnd:Swarm:ISSTA:2012}, the probability of inclusion of a test feature in the test program depends on the fair coin-toss for all test features. Therefore, the probability is the same for all test features, it is 0.5. 
Figure~\ref{fig:centroid:probabilities}, on the other hand, shows the distributions of probabilities per test feature in the \kconfig approach.
The mean and median of the probabilities of all features are less than 0.5. However, 28 out of 32 features have been in one or more configurations with probability 1.
In more than $25\%$ of configurations, $75\%$ or more of features are excluded.
It highlights the importance of a few test features that dominate the \tests generated by those clusters.

\section{Discussion}
\label{sec:discussion}

\kconfig and \swarm{} are complementary approaches in proposing configurations: \swarm{} chooses the configurations randomly, while \kconfig takes into account the bug reports in determining the configurations.
\kconfig generally finds more failures and provides higher code coverage than \swarm. Moreover, the number of distinct bugs detected by the techniques are: $4$ by both, $3$ only by \kconfig, and $3$ only by \swarm.

The particle swarm optimization technique used in \PSO requires a considerable amount of failing and passing \tests generated by \CS to find an optimal configuration. Unfortunately, as GCC matures, it becomes harder for \CS to trigger a failure in it---e.g., in Table \ref{table:result_10k}, compare the number of failures triggered by Default \CS in GCC 4.3.0 and 4.8.2. Therefore, \PSO loses its applicability to newer, more stable versions of GCC. Notice that in \cite{Junjie:PSO:ASE:2019}, \PSO is only trained on an older version (i.e. GCC-4.3.0), and in the cross-version experiment is tried in other versions (Section IV-F in \cite{Junjie:PSO:ASE:2019}).

\Part{Searching around the lamp posts}.
\KC analyzes features of failing \test{}s to create new \emph{configurations} for a test generator.
The result of our experiment shows that \KC could find up to \KMC{} miscompilations in GCC, and outperforms \swarm{} and \PSO in triggering miscompilation failures.

It suggests that analysis of the regression test suite can enable designing techniques to guide random test generators such as \CS. 
The result indicates that processing failing \tests can provide insights into the regions of code that are susceptible to bugs.
It might suggest that many bug fixes are incomplete~\cite{Yin:IncompleteBugFix:ESEC:FSE:2011}, but we are unable to verify the similarity between root causes of the bugs that \KC triggers and the \tests of regression test suite.

\Part{\regression vs. generated test suite}.
The coverage of GCC for the \tests in the regression test suite shows the power of small, directed \tests.
Although the size of the regression test suite was smaller than the test suites generated by \CS, they significantly outperformed those test suites in the coverage of GCC in terms of statements, branches, and functions.
The coverage of GCC for the regression test suite can serve as a benchmark to measure the shortcomings of the generated test suites.
From this experiment, the gap still is very wide. 
Note that we only could compile the \tests in the regression test suite without linking them because they did not accompany a \texttt{main} function.

\Part{Difference with deep learning-based approaches}.
This approach is different from (deep) learning-based approaches such as DeepSmith~\cite{Cummins:TestGenLearn:DeepSmith:ISSTA:2018} that build a generative model for the \tests of the programs. 
Moreover, learning-based techniques face two challenges. 
First, learning-based approaches require many \tests with millions of tokens to train a model. 
Second, learning-based approaches tend to converge to a restrictive language model of \test{} that overly restricts the type of \test{}s that can be produced ~\cite{Godefroid:FeatureBased:LearnFuzz:ASE:2017}. 
\KC instead uses the configuration of test generators to guide testing which is less constrained than the generation of \test{}s in learning-based approaches. 
In particular, \KC only specifies the programming constructs that should be present in the generated \test{}s, and the order or number of those constructs are determined by the test generator.

\Part{Limiting requirements}.
There are two main limitations to the application of \KC. First, it assumes that a stable test generator exists. Second, it requires a set of failing \tests. GCC compiler has been under development for decades and the bug reports are available.
Moreover, GCC has \CS \cite{Regehr:TestGenGrammar:Csmith:PLDI:2011}, a mature and well-engineered test generator, that allowed us to evaluate the effectiveness of \KC in testing GCC compilers.

\section{Related Work}
\label{sec:related}

Several approaches for testing compilers have been proposed; for example, 
code mutation \cite{Zeller:TestGenMutation:LangFuzz:Security:2012}, 
random test generation \cite{Regehr:TestSelection:Rank:PLDI:2013}, 
metamorphic testing \cite{Zhendong:TestOracle:classfuzz:PLDI:2016, Zhendong:TestGenMutation:Orion:PLDI:2014, Zhendong:TestGenMutation:Athena:OOPSLA:2015,Zhendong:TestGenMutation:Hermes:OOPSLA:2016},
learning-based testing \cite{Cummins:TestGenLearn:DeepSmith:ISSTA:2018}, 
and swarm testing \cite{Groce:TestGenRnd:Swarm:ISSTA:2012}, to name few. 
These approaches either generate \tests from scratch by grammar \cite{Regehr:TestGenGrammar:Csmith:PLDI:2011} and learning \cite{Cummins:TestGenLearn:DeepSmith:ISSTA:2018}, 
or they create new \tests by manipulating \cite{Zeller:TestGenMutation:LangFuzz:Security:2012} or transforming the existing \tests, e.g., \cite{Zhendong:TestGenMutation:Orion:PLDI:2014}.

Swarm testing \cite{Groce:TestGenRnd:Swarm:ISSTA:2012} randomly chooses a subset of features available to generate new test cases. 
The generated test cases are very diverse and the evaluation result shows that this approach outperforms \CS{}'s default configuration in both code coverage and crash bug finding. 
SPE \cite{Zhendong:TestSelection:SPE:PLDI:2017} where authors enumerate a set of programs with different variable usage patterns. 
The generated diverse test cases exploit different optimization and the evaluation result shows that the skeletal program enumeration has confirmed bugs in all tested compilers. 
Two more related studies in this area are LangFuzz \cite{Zeller:TestGenMutation:LangFuzz:Security:2012} and Learn\&Fuzz \cite{Godefroid:FeatureBased:LearnFuzz:ASE:2017}. 
The LangFuzz approach extracts code fragments from a given code sample that triggered past bugs and then apply random mutation within a pool of fragments to generate test inputs. 
On the other hand, the Learn\&Fuzz approach uses the learnt seq2seq model to automate the generation of an input grammar suitable for PDF objects using different sampling strategies. 
Both approaches have revealed several previously unknown bugs in popular compilers and interpreters.

Several approaches to accelerate the speed of test selection and triage have been proposed; for example, \cite{Regehr:TestSelection:Rank:PLDI:2013}, \cite{Chen:TestSelection:LET:ICSE:2017}, \cite{Chen:TestSelection:4steps:ICSE:2018}, \cite{Chen:TestSelection:COP:IEEE-TSE:2018}.
Chen et al. \cite{Regehr:TestSelection:Rank:PLDI:2013} evaluate the impact of several distance metrics on test case selection and prioritization.
Chen et al. \cite{Chen:TestSelection:LET:ICSE:2017} proposed LET where authors use machine learning to schedule the test inputs. This learning-to-test approach has two steps: learning and scheduling. In learning steps, LET extracts a set of features from the past bug triggering test cases and then trains a capability model to predict the bug triggering probability of the test programs, and trains another time model to predict the execution time of the test programs. In scheduling steps, LET ranks the target test programs based on the probability of bug triggering in unit time. The evaluation result shows that the scheduled test inputs significantly accelerate compiler testing.
Another example in this area is COP \cite{Chen:TestSelection:COP:IEEE-TSE:2018} where authors predict the coverage information of compilers for test inputs and prioritize test inputs by clustering them according to the predicted coverage information. The result shows that COP significantly outperforms state-of-the-art acceleration approaches in test acceleration.

\section{Threats to Validity}
\label{sec:threat}

In this section, we describe several threats to validity for our study.

\Part{Internal Validity.} 
Despite our best effort, some bugs may exist in the tools and scripts used in this paper.
However, we used the well-tested programs in the implementation and evaluation of \KC to reduce the chance of mistakes.
We have taken care to ensure that our results are unbiased, and have tried to eliminate the effects of variability by repeating the experiments multiple times.

\Part{External Validity.}
We evaluated the \KC approach on \CS and several GCC versions, therefore, the extent to which it generalizes to other compilers and test generators is quite unknown.
In this approach, the feature set is restricted to the ones that can be translated to \CS configuration options, therefore the feature space is also limited.
Additionally, \KC relies on the features that have triggered compiler bugs in the past, therefore the effectiveness may decrease in future GCC versions for the bug fixes.
Another potential issue can be that the bug reports to build the \regression test suite may not be representative of all GCC bugs and can impact the clusters.

\section{Conclusion}
\label{sec:conclusion}

This paper proposed \KC, a test configuration generation approach, that uses the code snippets in the bug reports to guide the test generation.
Given a \regression test suite and a configurable test generator, \KC computes configurations for the test generator.

We extensively evaluated \KC with GCC regression test suite and \CS{} on eight GCC versions. 
The results suggest that the configurations computed by \KC approach can help \CS{} to trigger more miscompilation failures than the state-of-the-art approaches. 
The results signify the benefits of analyzing bug reports in generation of new \tests.
Our source code for this work is publicly available at \url{https://github.com/mdrafiqulrabin/kconfig/}.

\section{Acknowledgement}
We thank Junjie Chen and Guancheng Wang for computing the number of distinct bugs and helping with the experiments related to HiCOND.

\bibliographystyle{acm}
\bibliography{refs}
\end{document}